\documentclass[aps,prl,floatfix,showpacs,twocolumn,tightenlines,superscriptaddress]{revtex4}

\usepackage{amsmath}
\usepackage{graphicx}
\usepackage{epsfig}

\newcommand{\ket}[1]{\mbox{$|#1\rangle$}}

\newcommand{\ketbra}[2]{\mbox{$|#1\rangle\langle #2|$}}
\newcommand{\op}[1]{\mbox{\boldmath $#1$}}
\begin{document}

\title{Generating optical nonlinearity using trapped atoms}

\author{Alexei~Gilchrist}
\email{alexei@physics.uq.edu.au}
\affiliation{Centre for Quantum Computer Technology and
Department of Physics, The University of Queensland,
St Lucia, QLD, 4072 Australia.}
\author{G.~J.~Milburn}
\affiliation{Centre for Quantum Computer Technology and
Department of Physics, The University of Queensland,
St Lucia, QLD, 4072 Australia.}
\affiliation{Department of Applied Mathematics and Theoretical Physics,
University of Cambridge, Cambridge,UK }
\author{W.~J.~ Munro}
\affiliation{Hewlett-Packard Laboratories, Filton Road, Stoke Gifford,
Bristol BS34 8QZ, United Kingdom}
\author{Kae~Nemoto}
\affiliation{School of Informatics, Dean Street, Bangor University,
Bangor LL57 1UT, United Kingdom}

\date{\today}

\begin{abstract}
  We describe a scheme for producing an optical
  nonlinearity using an interaction with one or more ancilla two-level
  atomic systems.  The nonlinearity, which can be implemented using
  high efficiency fluorescence shelving measurements, together with
  general linear transformations is sufficient for simulating
  arbitrary Hamiltonian evolution on a Fock state qudit. We give two
  examples of the application of this nonlinearity, one for the
  creation of nonlinear phase shifts on optical fields as required in
  single photon quantum computation schemes, and the other for the
  preparation of optical Schr\"odinger cat states.
\end{abstract}


\maketitle

Our ability to perform quantum transformations on optical fields is
hampered by the lack of materials with intrinsic optical
non-linearities.  While it is possible to circumvent this problem with
schemes conditioned on photo-detection (see for instance
\cite{KLM,99gottesman390,01pittman062311,02ralph012314,02knill052306,Kok2002}),
efficiency problems are encountered both due to the inherent nature of
the schemes and efficiencies of current photo-detectors. In this paper
we propose the use of the interaction with a simple atomic system,
conditioned on high efficiency atomic measurements, to generate a
near-deterministic (with probability approaching one) nonlinearity on
an optical field. This nonlinearity, together with linear
transformations, is sufficient for generating arbitrary Hamiltonian
evolution on a qudit formed from a truncated sequence of Fock states.

The interaction between the single mode field and a two-level atom is
described by the effective Hamiltonian $\op{H}=\kappa(\op{a}^\dagger
\op{\sigma}^-+\op{a}\op{\sigma}^+)$ where $\op{a}^\dagger$, $\op{a}$
are the boson creation and annihilation operators for the field mode
and the operators $\op{\sigma}^+=\ketbra{e}{g}$,
$\op{\sigma}^-=\ketbra{g}{e}$ are the raising and lowering operators
for the atomic state. The unitary transformation that acts when this
interaction is applied for a time $t$ is
$\op{U}(\tau)=\exp[-i\tau(\op{a}^\dagger
\op{\sigma}^-+\op{a}\op{\sigma}^+)]$ where $\tau=\kappa t$.

\begin{figure}
  \begin{center}
   \includegraphics[]{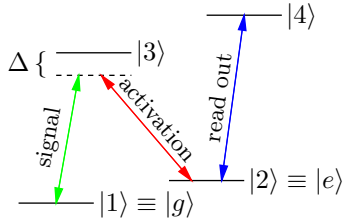}
  \end{center}
  \caption{Level scheme for an effective two-level transition controlled by a
    stimulated Raman process.}
  \label{fig1}
\end{figure}

If the atom is prepared in the ground state and found in the ground state
 after the interaction, the conditional state of the field is given by
\begin{equation}
\label{eq:a-dag-a}
\op{\Upsilon}_{g}(\tau)\ket{\psi} =
\cos(\tau\sqrt{\op{a}^\dagger\op{a}})\ket{\psi}
\end{equation}
On the other hand if the atom is prepared in the excited state and found in
the excited state after the interaction, the conditional state is given by
$\op{\Upsilon}_{e}(\tau)\ket{\psi} =
\cos(\tau\sqrt{\op{a}\op{a}^\dagger})\ket{\psi}$.
There is considerable practical advantage to using the excited state
preparation rather than the ground state as there is always a signal for
correct operation, however the analysis is the same, and we will concentrate on
the operator~(\ref{eq:a-dag-a}).

We have in mind a quantum computing communication protocol in which the
optical field mode is derived from a transform limited pulsed field which
is rapidly switched into the cavity mode containing the atomic systems at
fixed times determined by the pulse repetition rate.  Similar systems have
been proposed as a quantum memory for optical information processing
\cite{Pittman2002}.  When the atomic measurement yields the required result
the field may be switched out again for further analysis or subsequent
processing through linear and conditional elements.

Once the cavity field is prepared, we need to switch on the
interaction with the atomic system. In order that we can switch this
interaction at predetermined times we propose that an effective two
level transition connected by a Raman process with one classical field
and the quantised signal field, be used.  A similar scheme has
recently been proposed as the basis of a high efficiency photon
counting measurement \cite{Imamoglu,James_Kwiat}. The process is also
used in the EIT schemes for storing photonic information
\cite{Lukin2000} and for quantum state transfer between distant
cavities \cite{Cirac1997}. The level diagram is shown in figure
\ref{fig1}. The nearly degenerate levels $|1\rangle$ and $|2\rangle$
are connected by a stimulated Raman transition to level $|3\rangle$.
The detuning of the Raman pulse from the excited state $|3\rangle$ is
$\Delta$, which is approximately the same as the detuning of the
signal mode form the same transition. With these parameters the
interaction strength is given by $\kappa=\Omega
g/2\Delta$ where $\Omega$ is the Rabi frequency for the Raman pulse
and $g$ is the one photon Rabi frequency for the signal field.

An advantage of using a stimulated Raman process of this kind is that
the excited state $|2\rangle$ can be a meta-stable, long lived level.
We thus do not need to consider spontaneous emission from this level
back to the ground state.  The readout of the atomic system may be
achieved by using a cycling transition between the excited state
$|2\rangle$ and another probe level $|4\rangle$.  Such measurements
are routinely performed in ion trap studies \cite{Rowe2001} and can
have efficiencies greater than 99\%.

In the following, we will prove that the ability to perform the
conditional transformation~(\ref{eq:a-dag-a}) together with linear
transformations is universal on qudits. We will follow this with two
examples.

It is first instructive to consider a unitary operator of the form
$\exp(i\theta\sqrt{\op{a}^\dagger \op{a}})$.  A 
nonlinearity of the form $\sqrt{\op{a}^\dagger \op{a}}$, which we
refer to as the square-root-number operator, turns out to be as good
as a Kerr nonlinearity for universal quantum computation in the
infinite dimensional Hilbert space of a Harmonic oscillator
\cite{Braunstein99}.  The concept of universal computation is
isomorphic to the concept of universal simulation, which is the
ability to simulate any arbitrary Hamiltonian evolution, to any degree
of accuracy, by combining the evolution due to a fixed class of
Hamiltonian generators.

To motivate this idea, consider what kind of nonlinear oscillator
would correspond to the Hamiltonian $\op{H}_{\mathrm{srn}}=\sqrt{\op{a}^\dagger
  \op{a}}$. The resulting Heisenberg equation of motion for the field
amplitude operator is
\begin{equation}
\dot{\op{a}}=i[\op{H}_{\mathrm{srn}},\op{a}]=-i(\sqrt{\op{a}^\dagger \op{a}+1}-\sqrt{\op{a}^\dagger \op{a}})\op{a}
\end{equation}
which clearly indicates an oscillator in which the frequency is state
dependent.  The right hand side may be expanded as a power series in
$(\op{a}^\dagger \op{a})^{-1/2}$ to give
\begin{equation}
\dot{\op{a}}=-i\left(\frac{1}{2}(\op{a}^\dagger \op{a})^{-1/2}
-\frac{1}{8}(\op{a}^\dagger\op{a})^{-3/2}
+\ldots\right )\op{a}
\end{equation}
This corresponds to an oscillator for which the frequency of
oscillation {\em decreases} with increasing energy. For comparison, the
Kerr nonlinear oscillator corresponds to an oscillator in which the
frequency increases with increasing energy. It is thus clear that the
square-root-number Hamiltonian will result in a rotational shearing of
states in the phase plane of the oscillator in much the same way as
occurs for the Kerr nonlinearity \cite{Milburn86}.

We can go further by using the results of Braunstein and Lloyd
\cite{Braunstein99}.  They considered the question of what
Hamiltonians are universal for quantum computation in an infinite
dimensional Hilbert space, such as that for a single mode of the
radiation field. Their results show that Hamiltonians that are at most
quadratic in the canonical momentum and position variables are not
universal.  For instance, in the case of a single mode field with
annihilation and creation operators $\op{a},\op{a}^\dagger$, successive
applications of Hamiltonians from the set of displacements, squeezing
and rotations, $\mathcal{H}_{\mathrm{lin}}=\{z\op{a}+z^*\op{a}^\dagger,
  z\op{a}^2+z^*(\op{a}^\dagger)^2, z\op{a}^\dagger \op{a}\}$, can
generate arbitrary linear canonical transformations in the phase space
variables but no other transformations.  A universal set is easily
obtained by adjoining almost any Hamiltonian that is at least cubic in
the canonical variables. A universal set of Hamiltonians could be made
up from $\mathcal{H}_{\mathrm{lin}}$ together with the Kerr nonlinearity
$\op{H}_k=(\op{a}^\dagger)^2 \op{a}^2$. Another choice is the cubic
Hamiltonian $\op{H}_c=\op{a}^\dagger
\op{a}(\op{a}+\op{a}^\dagger)+hc$.  We now show that the
square-root-number operator $\sqrt{\op{a}^\dagger \op{a}}$ together
with the set $\mathcal{H}_{\mathrm{lin}}$ can be used to simulate a cubic
Hamiltonian, and thus is universal for quantum simulations in the
Hilbert space of a single mode.

Consider acting on the unitary operator generated by the square-root-number
operator with a large displacement $\alpha$:
$\op{U}(\alpha,\theta)=\op{D}^\dagger(\alpha)\exp(i\theta\sqrt{\op{a}^\dagger \op{a}})\op{D}(\alpha)$
where $\op{D}(\alpha)$ is the displacement operator given by 
$\op{D}(\alpha)=\exp(\alpha \op{a}^\dagger -\alpha^* \op{a})$. The resulting unitary
operator may then be written as
\begin{eqnarray}
\op{U}(\alpha,\theta)&=&e^{i\theta|\alpha|}\exp\left[i\theta\left(
\frac{\op{x}(\phi)}{2}+\frac{\op{a}^\dagger\op{a}}{2|\alpha|}
-\frac{\op{x}(\phi)^2}{8|\alpha|} \right.\right.\nonumber\\
&&\left.\left.-\frac{\op{a}^\dagger \op{a} \op{x}(\phi)+\op{x}(\phi) \op{a}^\dagger\op{a}}{8|\alpha|^2}+ O(|\alpha|^{-3})\right)\right]
\end{eqnarray}
where $\op{x}(\phi)=(\op{a}e^{-i\phi}+hc)$ with $\phi$ the phase of
$\alpha$. The first three terms in the argument of the exponential
correspond to a second order Hamiltonian and simulate a displacement,
a rotation, and a squeezing operation respectively. The fourth term
however is a cubic term which is what we required for a universal set
of Hamiltonians for a single mode. By using a linear Hamiltonian to
mix several modes (for instance a beam-splitter), arbitrary multi-mode
Hamiltonians can be constructed \cite{Braunstein99}.  It is thus clear
that we can use the square-root-number operator, together with an
arbitrary linear transformations to perform universal computation.

Now that we have shown the universality of the operator $\exp(i \theta
\sqrt{\op{a}^\dagger \op{a}})$ what can we say directly about $\cos
\left[ \theta \sqrt{\op{a}^\dagger \op{a}}\right]$? The action of this
conditional operator on the number state $\ket{n}$ is to multiply the
state by the amplitude
$\cos(\theta\sqrt{n})=[\exp(i\theta\sqrt{n})+\exp(-i\theta\sqrt{n})]/2$.
Clearly, if we can restrict the interaction so that this amplitude is
$\pm 1$, then it is equivalent to the full unitary operator---and we
might expect to exploit the nonlinear interaction in a similar way.
Now, as it turns out, it \emph{is} possible \footnote{With
  $\cos(\theta\sqrt{2})=-1$, the fundamental theorem of arithmetic
  implies that for $\sqrt{n}$ which are rational multiples of
  $\sqrt{2}$, $\cos(\theta\sqrt{n})=\pm 1$. Irrational multiples of
  $\sqrt{2}$ will have incommensurate periods, so it should be possible
  to find $\theta$ for which (\ref{qudit:rel}) is true to within a
  given error.}  to choose $\theta$ such that for a finite size
computational space, $\ket{0}\ldots\ket{N}$,
\begin{equation}\label{qudit:rel}
\cos \left[ \theta \sqrt{\op{a}^\dagger \op{a}}\right]\ket{n} \approx
\begin{cases}
  -\ket{n}& \text{for $n=2 \left(2 m +1\right)^2$}\\
  \ket{n} & \text{for $n\neq2 \left(2 m +1\right)^2$}
\end{cases}
\end{equation}
Here $m$ is a non-negative integer and we see a sign shift only on the
states   $|2\rangle, |18\rangle, |50\rangle \ldots | 2 \left(2 m
+1\right)^2\rangle \ldots$. We have chosen the first sign shift to
occur on the $|2\rangle$ Fock state. This is an arbitrary choice and
other terms can be shifted instead.

For most computation schemes in which we are interested, there is a
finite number of basis states (there generally always is a natural
cutoff even in most CV schemes). In this case it is always possible to
choose $\theta$ to satisfy the above relations. This indicates that
our operator $\cos \left[ \theta \sqrt{\op{a}^\dagger \op{a}}\right]$
is approximating an effective, highly nonlinear unitary operator.  For
instance, if we consider only the qutrit subspace of $\ket{0}$,
$\ket{1}$, and $\ket{2}$ then the operator acts equivalently to the
Hamiltonian, $H=\frac{\pi}{2}\op{a}^{\dagger 2} \op{a}^2$ which
contains a Kerr nonlinearity.

With this nonlinear Hamiltonian, it is possible in conjunction linear
Hamiltonians to generate arbitrary multi-mode Hamiltonians. Hence we
can in principle create any of the unitary operators required for
universal qutrit computation. This argument extends to higher qudits
in a direct manner.  The potential problem however, is that the
greater the number of states in the qudit space, the larger the value
of $\theta$ required to satisfy (\ref{qudit:rel}).

It has recently been shown by Knill, Laflamme and Milburn (KLM)
\cite{KLM}, that a conditional nonlinear sign shift (NS)
gate on two photon states can be produced with passive linear optical
elements and photo-detection.  Such conditional nonlinear phase shifts
can be used to perform two qubit operations for logical states encoded
in photon number states.  If such conditional gates are used to
prepare entangled states for teleportation, efficient quantum
computation can be performed which with suitable error correcting
codes can be made fault tolerant \cite{KLM}.  Here we show that if the
ancilla modes are replaced with a two level atom similar conditional
nonlinear phase shifts can be achieved by near deterministic
post-selection on atomic measurements.  The atomic measurements can be
made with fluorescence shelving techniques which are very much more
efficient than single photon counting measurements, thus reducing the
need for new photon counting technologies inherent in the KLM scheme.
In addition, the near-deterministic nature of our gate would also
drastically simplify the implementation of the KLM scheme.  The
ability to induce conditional nonlinear phase shifts on quantum
optical fields also has applications to high precision measurements
(see for instance \cite{Kok2002}).

Our objective for the NS gate is to find a way to produce the
nonlinear phase shift defined by
\begin{equation}\label{eq:ns}
c_0\ket{0}+c_1\ket{1}+c_2\ket{2}\rightarrow 
c_0\ket{0}+c_1\ket{1}-c_2\ket{2}
\end{equation}
where the $c_j$ are complex amplitudes satisfying $\sum |c_j|^2=1$.
This is in the form of the conditions in equation~(\ref{qudit:rel}),
so consider applying the measurement operator $\op{\Upsilon}_{g}$, on the
general two photon state. This will leave us with the state
$A_0c_0\ket{0}+A_1c_1\ket{1}+A_2c_2\ket{2}$
where $A_0=1$, $A_1=\cos(\tau)$ and $A_2=\cos(\sqrt{2}\tau)$. 
Solutions for which $(A_1,A_2)$
approaches arbitrarily close to $(1,-1)$ can 
easily found, for examples see table~\ref{tab:Yg}.
\begin{table}[htbp]
\begin{tabular}[c]{|c|ccc|}
\hline
$\tau$ & $A_0$ & $A_1$ & $A_2$ \\
\hline
6.5064 & 1 & 0.97519 & -0.97516 \\
37.73742 & 1 & 0.9992663 & -0.9992665 \\
219.918 & 1 & 0.999979 & -0.999978 \\
\hline
\end{tabular}\centering
\caption{High-probability results for a single atom initially in
a ground state and measured in a ground state after an interaction time
$\tau$. 
}
\label{tab:Yg}
\end{table}
It is also possible to employ several atoms, each addressed
independently, and find solutions of high probability using the same
techniques. For example with two atoms, one initially prepared in the
ground state and the second prepared in the excited state, the
conditional state given that both atoms are found in their initial
state after the interaction is
$\ket{\phi_{ge}}=\op{\Upsilon}_{g}(\tau_1)
\op{\Upsilon}_{e}(\tau_2)\ket{\psi}$ and again values of the
interaction times $\tau_1$ and $\tau_2$ can be found which again
perform the NS gate with high probability. For instance with
$\tau_1=37.79300921$ and $\tau_2=197.78109842$, then
$|A_1|=|A_2|=|A_3|=0.990321935$ and the required phase shift is
performed on $\ket{2}$.

In any real experiment, the desired interaction time can be calibrated by
placing the phase shift in one arm of a Mach-Zehnder interferometer,
with two single photon inputs, and examining the interference fringes as a
function of the interaction time.

\begin{figure}
  \begin{center}
   \includegraphics[width=1.0\columnwidth]{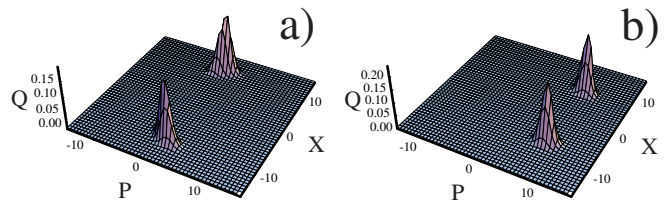}
  \end{center}
  \caption{A plot of the Q-function versus the two canonical phase
    space variables, for conditional state produced from an initial
    coherent state using, (a) $\alpha=10,\;\theta=10\pi$ and (b)
    $\alpha=10,\;\theta=5\pi$}
  \label{fig3}
\end{figure}
Now let us consider the case of the field in an initial coherent state
$|\alpha\rangle$. We will show that the conditional transformation,
$\cos(\theta\sqrt{\op{a}^\dagger \op{a}})$, generates a coherent
superposition of coherent states, a so called {\it Schr\"odinger cat state}.
A convenient phase space representation of the conditional state is the Q-function
defined by
\begin{eqnarray}
Q(\beta)&=&|\langle\beta|\cos(\theta\sqrt{\op{a}^\dagger \op{a}})|\alpha\rangle|^2 \nonumber \\
&=&\frac{1}{4}|\langle\beta|e^{i \theta\sqrt{\op{a}^\dagger \op{a}}}+e^{-i \theta\sqrt{\op{a}^\dagger \op{a}}}|\alpha\rangle|^2 \nonumber
\label{exact}
\end{eqnarray}
We thus first consider the amplitude function
${\cal A}(\beta)=\langle \beta|\exp(i\theta\sqrt{\op{a}^\dagger \op{a}})|\alpha\rangle$.
Based on the semi-classical expectation that this unitary
transformation describes an oscillator with an energy dependent
frequency, we anticipate that we need only consider the Q-function on
the curve $|\beta|=|\alpha|$. With this in mind we put
$\beta=|\alpha|e^{i\phi}$ and the Q-function amplitude is then given
by
\begin{equation}
 {\cal A}(\beta)=\sum_{n=0}^\infty p_n(\alpha) e^{i(\theta\sqrt{n}-\phi n)}
\end{equation}
where $p_n(\alpha)=e^{-|\alpha|^2}\frac{|\alpha|^n}{n!}$.  We now
assume that $|\alpha|>>1$ and approximate the Poisson distribution,
$p_n(\alpha)$ with a Gaussian
\begin{equation}
p_n(\alpha)\approx(2\pi|\alpha|)^{-1/2}\exp\left
[-\frac{(n-|\alpha|)^2}{2|\alpha|}\right ].
\end{equation}
We can then replace the sum with an integral over the variable
$y=n-|\alpha|$. Under the assumption
$|\alpha|>>1$ we find the integrand can be approximated as a general
Gaussian and thus the integral is
given by
\begin{equation}
  {\cal A}(\beta)=e^{-i\phi|\alpha|^2+i\theta|\alpha|}\exp\left
[-\frac{1}{8}(\theta-2|\alpha|\phi)^2\right ].
\end{equation}
Clearly this distribution is peaked at $\phi=\theta/(2|\alpha|)$. If
we choose the interaction time so that $\theta=|\alpha|\pi$, we expect
the state to be localised on the positive imaginary axis in the phase
plane of the Q-function.  Similarly for the same parameters $\langle
\beta|\exp(i\theta\sqrt{\op{a}^\dagger \op{a}})|\alpha\rangle$ will be
localised on the negative imaginary axis in the phase plane of the
Q-function.  It then follows that for the full conditional operator,
$\cos(\theta\sqrt{\op{a}^\dagger \op{a}})$, the state has two
components localised symmetrically about the origin on the imaginary
axis. In figure [\ref{fig3}] we show the Q~function as a function of
the two canonical phase space variables $x$ and $p$ for several choices of
$\alpha$ and $\theta$.

We observe that if we load a coherent state into the cavity, and
repeat the conditioning measurement procedure outlined previously,
the conditional state of the field will be prepared in a state which
is close to a cat state. Note however that these cats are not parity
eigenstates as the conditional interaction cannot change the photon
number distribution. Similar cat states are produced by a Kerr
nonlinearity \cite{cat1,cat2}.

We now estimate some typical values for the parameters. In a recent
experiment a similar stimulated Raman process was observed using single
rubidium atoms falling through a high finesse optical cavity
\cite{Henrich2000}. The following parameters are typical of that
experiment: $g=2\pi\times 4.5$~MHz, $\Omega=2\pi\times 30 $~MHz and
$\Delta=2\pi\times 6$~MHz. This gives a coupling constant of the order of
$70$~Mhz.  To achieve effective interaction constants of the order of those
in the table \ref{tab:Yg}, requires interaction times of the order of
$0.1-5$~$\mu$s.  In this paper we have neglected cavity decay which
obviously needs to be kept small over similar time scale, which while
difficult is not impossible.

We have shown how the ability to do very efficient measurements on
single atoms trapped in an optical cavity can be used to implement
nonlinear conditional phase shifts on the intra-cavity field. By
carefully choosing the interaction time, a nonlinear interaction can
be implemented with near unit probability, that, together with linear
transformations, is universal for simulating an interaction on qudits.
For small qudits the required interaction time can easily be found
numerically.  If the field state can be carefully switched in and out
of the cavity, the method can be used to implement near-deterministic
nonlinear gates for quantum optical computing. For instance, the
scheme can be used to implement a nonlinear sign shift gate, which
thus provides a path to quantum computation with logical qubits
encoded in photon number states. It can also be used to conditionally
generate coherent superpositions of coherent states and thus can
provide the key resource for quantum computing with coherent states
\cite{Ralph02}.

We would like to thank the Computer Science department of the
University of Waikato for making available their computing resources.
AG was supported by the New Zealand Foundation for Research, Science
and Technology under grant UQSL0001. GJM was supported by the
Cambridge-MIT Institute while a visitor at University of Cambridge.
WJM acknowledges support from the EU through the project RAMBOQ.
KN acknowledges support from the Japanese Research Foundation for
Opto-Science and Technology.
%=====================================================================

%\bibliographystyle{prsty}
%\bibliography{references}

\begin{thebibliography}{10}

\bibitem{KLM}
{E. Knill}, {R. Laflamme}, and {G. Milburn}, Nature {\bf 409},  46  (2001).

\bibitem{99gottesman390}
D. Gottesman and I.~L. Chuang, Nature {\bf 402},  390  (1999).

\bibitem{01pittman062311}
T.~B. Pittman, B.~C. Jacobs, and J.~D. Franson, Phys. Rev. A {\bf 64},  062311
  (2001).

\bibitem{02ralph012314}
T.~C. Ralph, A.~G. White, W.~J. Munro, and G.~J. Milburn, Phys. Rev. A {\bf
  65},  012314  (2002).

\bibitem{02knill052306}
E. Knill, Phys. Rev. A {\bf 66},  052306  (2002).

\bibitem{Kok2002}
P. Kok, H. Lee, and J.~P. Dowling, Phys. Rev. A {\bf 65},  052104  (2002).

\bibitem{Pittman2002}
T.~B. Pittman and J.~D. Franson, Cyclical quantum memory for photonic qubits,
  2002.

\bibitem{Imamoglu}
A. Imamo\"glu, Phys. Rev. Lett. {\bf 89},  163602  (2002).

\bibitem{James_Kwiat}
D.~F.~V. James and P.~G. Kwiat, Phys. Rev. Lett. {\bf 89},  183601  (2002).

\bibitem{Lukin2000}
M. Fleischhauer and M.~D. Lukin, Phys. Rev. Lett. {\bf 84},  5094  (2000).

\bibitem{Cirac1997}
J.~I. Cirac, P. Zoller, H.~J. Kimble, and M. Mabuchi, Phys. Rev. Lett.  3221
  (1997).

\bibitem{Rowe2001}
M.~A. Rowe {\it et~al.}, Nature {\bf 409},  791  (2001).

\bibitem{Braunstein99}
S. Lloyd and S.~L. Braunstein, Phys. Rev. Lett. {\bf 82},  1784  (1999).

\bibitem{Milburn86}
G.~J. Milburn, Phys. Rev. A {\bf 33},  674  (1986).

\bibitem{cat1}
V. Buzek and P.~L. Knight,  in {\em Progress in Optics}, edited by E. Wolf
  (Elsevier, Amsterdam, 1995).

\bibitem{cat2}
C.~C. Gerry and P.~L. Knight, Am. J. Phys. {\bf 65},  973  (1997).

\bibitem{Henrich2000}
M. Henrich, T. Legero, A. Kuhn, and G. Rempe, Phys. Rev. Lett. {\bf 85},  4872
  (2000).

\bibitem{Ralph02}
T.~C. Ralph, W.~J. Munro, and G.~J. Milburn, Quantum Computation with Coherent
  States, Linear Interactions and Superposed Resources., 2001,
  quant-ph/0110115.

\end{thebibliography}

\end{document}